\documentclass[]{ceurart}

\usepackage{listings}
\usepackage[capitalise, noabbrev]{cleveref}
\usepackage{booktabs}
\usepackage{tabularx}
\usepackage{xurl}
\usepackage{xspace}
\newcommand{\codename}{INSIGHT\xspace}
\begin{document}

\copyrightyear{2025}
\copyrightclause{Copyright for this paper by its authors. Use permitted under Creative Commons License Attribution 4.0 International (CC BY 4.0).}

\conference{D-SAIL Workshop - Transformative Curriculum Design: Digitalisation, Sustainability, and AI Literacy for 21st Century Learning, July 22, 2025, Palermo, Italy}

\title{INSIGHT: Bridging the Student-Teacher Gap in Times of Large Language Models}

\author{Jarne Thys}[
orcid=0009-0005-9348-7405,
email=jarne.thys@uhasselt.be,
url=https://www.uhasselt.be/en/who-is-who/jarne-thys,
]
\address{UHasselt -- Hasselt University, Digital Future Lab -- Flanders Make, Wetenschapspark 2, 3590 Diepenbeek, Belgium}
\cormark[1]

\author{Sebe Vanbrabant}[
orcid=0009-0001-7996-6048,
email=sebe.vanbrabant@uhasselt.be,
url=https://www.uhasselt.be/en/who-is-who/sebe-vanbrabant,
]

\author{Davy Vanacken}[
orcid=0000-0001-8436-5119,
email=davy.vanacken@uhasselt.be,
url=https://www.uhasselt.be/en/who-is-who/davy-vanacken,
]

\author{Gustavo {Rovelo Ruiz}}[
orcid=0000-0001-7580-8950,
email=gustavo.roveloruiz@uhasselt.be,
url=https://www.uhasselt.be/en/who-is-who/detail/gustavo-roveloruiz,
]

\cortext[1]{Corresponding author.}

\begin{abstract}
The rise of AI, especially Large Language Models, presents challenges and opportunities to integrate such technology into the classroom. AI has the potential to revolutionize education by helping teaching staff with various tasks, such as personalizing their teaching methods, but it also raises concerns, for example, about the degradation of student-teacher interactions and user privacy. Based on interviews with teaching staff, this paper introduces \codename, a proof of concept to combine various AI tools to assist teaching staff and students in the process of solving exercises. \codename has a modular design that allows it to be integrated into various higher education courses. We analyze students' questions to an LLM by extracting keywords, which we use to dynamically build an FAQ from students' questions and provide new insights for the teaching staff to use for more personalized face-to-face support. Future work could build upon \codename by using the collected data to provide adaptive learning and adjust content based on student progress and learning styles to offer a more interactive and inclusive learning experience.
\end{abstract}


\maketitle

\section{Introduction}
Due to the rising popularity of Artificial Intelligence (AI), various educational tools incorporating AI have been developed in recent years. AI can provide personalized learning experiences by tailoring educational experiences to individual learners' needs and abilities~\cite{chen_artificial_2020}. For example, adapting content and pace can improve learning effectiveness and efficiency~\cite{shemshack_systematic_2020,yu_personalized_2020}. AI enables personalized learning at any time, making education more efficient and accessible to larger groups~\cite{altinay_factors_2024}. Although most research focuses on K-12 settings, personalized learning is also promising for higher education, which is the main focus of this work, as AI can consider the students' unique motivations and circumstances~\cite{yu_personalized_2020}. Recent research also focuses on making Large Language Models (LLMs) fit for the classroom by reducing hallucinations, which is achieved by using course materials as input for the answer~\cite{olney_jill_2024}. In mathematics education, LLMs show promise in developing students' problem-solving skills and critical thinking abilities by enabling them to engage more deeply with open-ended problems through iterative questioning, clarification of concepts, and exploration of multiple solution paths~\cite{trotter_ai-assisted_2024}. Nitze~\cite{nitze_future-proofing_2024} developed a prototype to simulate oral exams in STEM education to reduce educator workload and give students individual feedback in the early stages of their academic journey.

Despite the numerous benefits AI offers to students and teaching staff, its integration into education raises several concerns. First, AI can degrade human interaction in education~\cite{amanda_puteri_impact_2024,owan_exploring_2023}, as it reduces face-to-face engagement and creates excessive dependence on the technology~\cite{altinay_factors_2024,kasneci_chatgpt_2023}, which could lead to worse educational experiences for students. Second, as students start asking more questions to LLMs, the teaching staff might miss important cues, such as gaps in comprehension of the learning material~\cite{kasneci_chatgpt_2023,owan_exploring_2023}. Third, privacy is an important concern, as AI tools collect a vast amount of data. OpenAI, for example, collects the user's input content (including files), device information, and location information~\cite{openai_europe_2024}. Educational applications, when deployed into a classroom, collect additional data that needs to be stored in a way that is both ethical~\cite{huang_ethics_2023} and compliant with applicable laws, such as the EU's GDPR. Anthology's BlackBoard, for example, collects, among other data, a student's solutions, grades, progress on a per-file basis, average hours in the course, and days of inactivity~\cite{virginia_wesleyan_university_student_nodate}.

To address these challenges, we introduce \codename: INtelligent Student and Instructor Guidance through Human-Centered Technology, which implements different AI tools to support \textbf{student-teacher interaction} in a privacy-aware manner. Our contribution is twofold. First, \codename equips teaching staff with \textbf{data-driven insights} on students' questions and challenges regarding a specific course topic. Based on these insights, they can support each student in a tailored manner and adapt their course material based on the needs of students. Second, \codename provides students access to an LLM in a monitored environment while ensuring \textbf{data privacy} through explicit opt-in data collection.

\section{Interviews with teaching staff}
To guide \codename's development, we organized focus groups with various teaching staff at Hasselt University. This was done in two phases. In the first phase, we conducted semi-structured interviews with members of the teaching staff from different courses. An initial prototype was built using this feedback. Afterward, we collected additional input by demonstrating the prototype at an internal workshop that specifically targeted AI in the university. This way, we could get input from a variety of people with different backgrounds.

The main takeaway from these interactions is that teaching staff are concerned they might lose contact with students, and thus, hinder their insights into the comprehension levels of their course knowledge. Traditionally, they could determine this based on how many and what kind of questions students would ask them. With the emergence of tools such as ChatGPT, they are concerned that students will start asking most of their questions to those kinds of tools, leaving the teaching staff with little feedback to improve their course. This was particularly relevant for a subset of the teaching staff who had only recently developed a new master's program with completely new courses and course material.

\section{\codename's modular framework for AI-powered student-teacher interaction}
\begin{figure}
    \centering
    \includegraphics[width=0.6\linewidth]{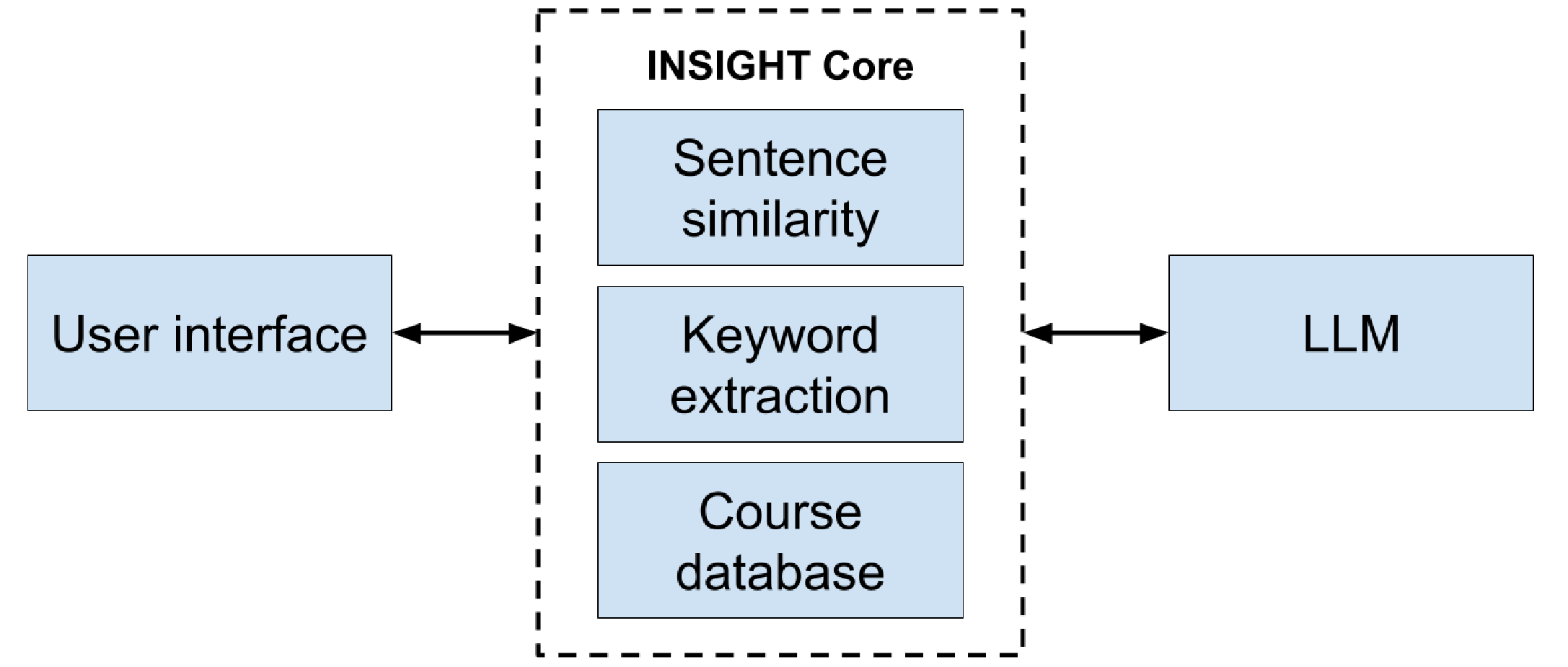}
    \caption{\codename has a central system, \textit{\codename Core}, with a keyword extraction and sentence similarity model and a database containing all course information and student interactions with \codename while solving exercises. \codename Core is connected to a local LLM and a user interface that can be independently changed based on specific needs.}
    \label{fig:arch}
\end{figure}

\codename's components are designed modularly, enabling \codename to be deployed across various domains without changing its core implementation. We achieve this flexibility by dividing the solution into multiple independently adjustable components, as illustrated in \cref{fig:arch}. We have a central system, \textit{\codename Core}, with a keyword extraction and sentence similarity model and a database. This database contains all the course information and data on student interactions with \codename while solving exercises (e.g., questions to the LLM, FAQ usage, and difficulty ratings). \codename Core is connected to a local LLM and a user interface that can be independently changed based on specific needs. Importantly, \codename does not interfere with the behavior of the LLM in any way. There is no prompt engineering, rephrasing, or nudging of responses, ensuring students interact with it as they naturally would and are not discouraged by artificial limits. As a proof of concept, we incorporate data from Hasselt University's Algorithms and data structures course~\cite{uhasselt_algoritmen_2023} to demonstrate and assess \codename's feasibility. An overview of \codename's features and their benefits for teaching staff and students can be found in \cref{tbl:overview}.

\begin{table}
\centering
\begin{tabularx}{\linewidth}{p{0.2\linewidth} p{0.28\linewidth} X}
\toprule
\multicolumn{1}{c}{\textbf{Feature}} & \multicolumn{1}{c}{\textbf{Benefit for teaching staff}}  & \multicolumn{1}{c}{\textbf{Benefit for students}}                                      \\ \midrule
LLM chat interface                     & Reduction in workload as it can answer simpler questions & Easy access to answers, just like using other LLM tools                                \\
Dynamic FAQ                            & Allow verification of the answer to common questions     & Have a teacher-verified answer for common questions                                    \\
Teaching insights                      & Quick indication of topics of recurring questions        & Get more personalized feedback, as the teacher knows where they might struggle         \\
Data anonymization                     & /                                                        & Be in control of their data and have the option to share specific data they agree to \\ \bottomrule
\end{tabularx}
\caption{Overview of the four \codename features --- an LLM‐based chat interface, a dynamic FAQ generator, data‐driven teaching insights, and a configurable data‐anonymization layer --- and the way each feature simultaneously benefits teaching staff (by reducing repetitive question load, enabling verification of common queries, and surfacing areas of recurring difficulty) and students (by providing instant, teacher‐reviewed answers, more personalized feedback in the classroom, and the option to share only the data they consent to).}
\label{tbl:overview}
\end{table}

\paragraph{\textbf{Answering student questions through LLMs.}}
LLMs have become increasingly widespread, which presents a valuable opportunity for the educational sector to provide frameworks that guide students in making the most of these technologies in a meaningful, ethical, and educationally responsible manner. We use an LLM as the main interaction method between students and \codename. We later analyze students' interactions with the LLM to provide the teaching staff with data they can use to offer better face-to-face support and increase student-teacher interaction. In this prototype, we used Llama 3.2 3B as it has an above-average output speed and below-average latency, ensuring quick responses~\cite{artificial_analysis_llama_nodate}. This comes at the cost of intelligence, but in our initial testing, it seems adequate for the use case. The small amount of parameters also makes it easier to deploy on consumer hardware. Finally, \codename's modular architecture allows us to easily switch to other (newer) models and pick the right model depending on the needs of the course. Gemma 3~\cite{gemma_team_gemma_2025} is an example of a recent development that could be interesting for our use case. At the time of writing, it a a state-of-the-art model optimized to run on a single GPU. This would make it feasible to, for example, run one high-performance PC locally in the classroom.

\paragraph{\textbf{Keyword mapping and dynamically generated FAQ from student questions.}}
After the teaching staff review the keywords for each exercise, \codename treats that list as the topic vocabulary for the course. All questions from students are embedded in the same vector space. The nearest keyword determines the question’s topic label. After identification, the topic is added to the analytics for the teaching staff. These student questions to the LLM can also help teaching staff identify challenging topics and knowledge gaps. Recurrent questions can be added to an FAQ, which is primarily designed as a tool for the teaching staff to monitor learning trends and gaps, but also serves as a resource for students facing similar issues. This new information also allows teaching staff to provide better face-to-face support for specific students or groups. To group similar questions in the FAQ, \codename uses the vector embeddings from the all-MiniLM-L6-v2 model to group semantically related questions based on their cosine similarity. This model was chosen because it was trained on multiple datasets, making it very generalizable, while it maintains a relatively high accuracy and speed~\cite{sbert_pretrained_nodate}. We note that this approach was developed for the given algorithms and data structures and might require fine-tuning to apply it to other courses. Another benefit of grouping similar questions is the ability to return the same answer to these questions. \codename caches the LLM's answer and provides the teaching staff the ability to edit the answer to their liking, ensuring human-in-the-loop involvement and reducing the variability of LLM answers.

\begin{figure}
    \centering
    \includegraphics[width=0.75\linewidth]{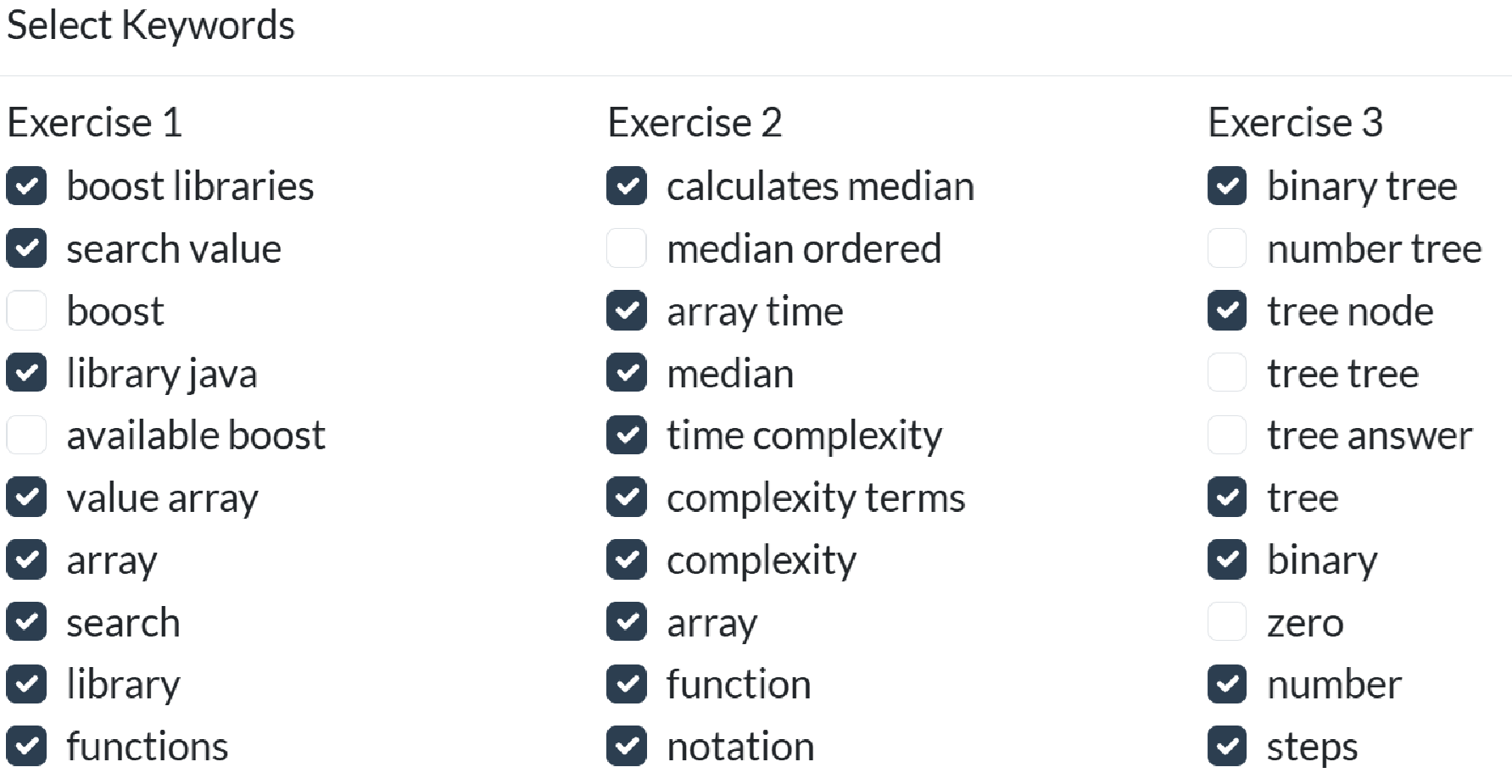}
    \caption{The keyword selection is a mixed-initiative process where the initial list of keywords is determined through a keyword extraction algorithm. The teaching staff can further refine the list for final use. \cref{tbl:keywords} shows the full texts for these three exercises, together with the automatically extracted keywords.}
    \label{fig:key-sel}
\end{figure}

\paragraph{\textbf{Leveraging student interaction data for teaching insights.}}
When students engage with an LLM, teaching staff may lose sight of their understanding of the course material, as the amount and nature of questions serve as a form of implicit feedback on the learning resources. This makes it harder for the teaching staff to provide support and refine those learning resources. To address this, we propose a system that collects and analyzes students' questions to the LLM, their FAQ usage, and explicit input in the form of difficulty ratings for the exercises they complete. This data provides valuable insights that, over time, can help refine course materials to better align with student needs and expectations. To determine the topics from students' questions to the LLM, \codename must first identify the topics covered in the exercises. To achieve this, we use keyword extraction to infer the exercise topics from the exercise descriptions. The keyword extraction is a mixed-initiative process that first uses KeyBERT~\cite{grootendorst_maartengrkeybert_2023} to extract an initial list of keywords from the exercise text. After extraction, the teaching staff can manually review the keywords for each exercise to best represent the tasks and make them suitable for identifying question topics. An example is illustrated in \cref{fig:key-sel}.

\begin{table}
\centering
\begin{tabularx}{\linewidth}{X X}
\toprule
\multicolumn{1}{c}{\textbf{Exercise text}} & \multicolumn{1}{c}{\textbf{Extracted keywords}} \\ \midrule
According to the documentation, which kind of functions do the C++ standard library and java provide to search a value in an array? What is available in the boost C++ libraries? Use these libraries (unless instructed otherwise) to make your work easier! & boost libraries, search value, boost, library java, available boost, value array, array, search, library, functions \\
This simple function calculates the median from an ordered array. Describe its time complexity in terms of the big o notation. What is the main difference here, compared to similar exercises on arrays when considering n? & calculates median, median ordered, array time, median, time complexity, complexity terms, complexity, array, function, notation \\
A binary tree is a tree in which each node has zero, one, or two children (see figure). How many steps would a search for a number take in such a tree? Answer in terms of n. Note that trees will be covered in later chapters. & binary tree, number tree, tree node, tree tree, tree answer, tree, binary, zero, number, steps \\ \bottomrule
\end{tabularx}
\caption{Extracted keywords for three exercises from the Algorithms and data structures course. Through the mixed-initiative approach, the teaching staff can manually refine the keywords, as the AI model can make mistakes. An example can be seen in Exercise 3, where the model recommends 'tree tree' as a good keyword.}
\label{tbl:keywords}
\end{table}

\paragraph{\textbf{Data anonymization policy.}}
While it can be helpful for teaching staff to identify exercises and topics students find challenging, not all students might be comfortable sharing identifying data. To address this, students can select the \textit{anonymous mode}, allowing teaching staff access to the data without the students' names attached. Additionally, \codename uses Ollama to run an LLM locally, ensuring student queries, potentially containing sensitive information, remain private and are never shared with third parties. This method ensures that personal data never leaves the students' devices without their explicit consent.

\begin{figure}
    \centering
    \includegraphics[width=0.75\linewidth]{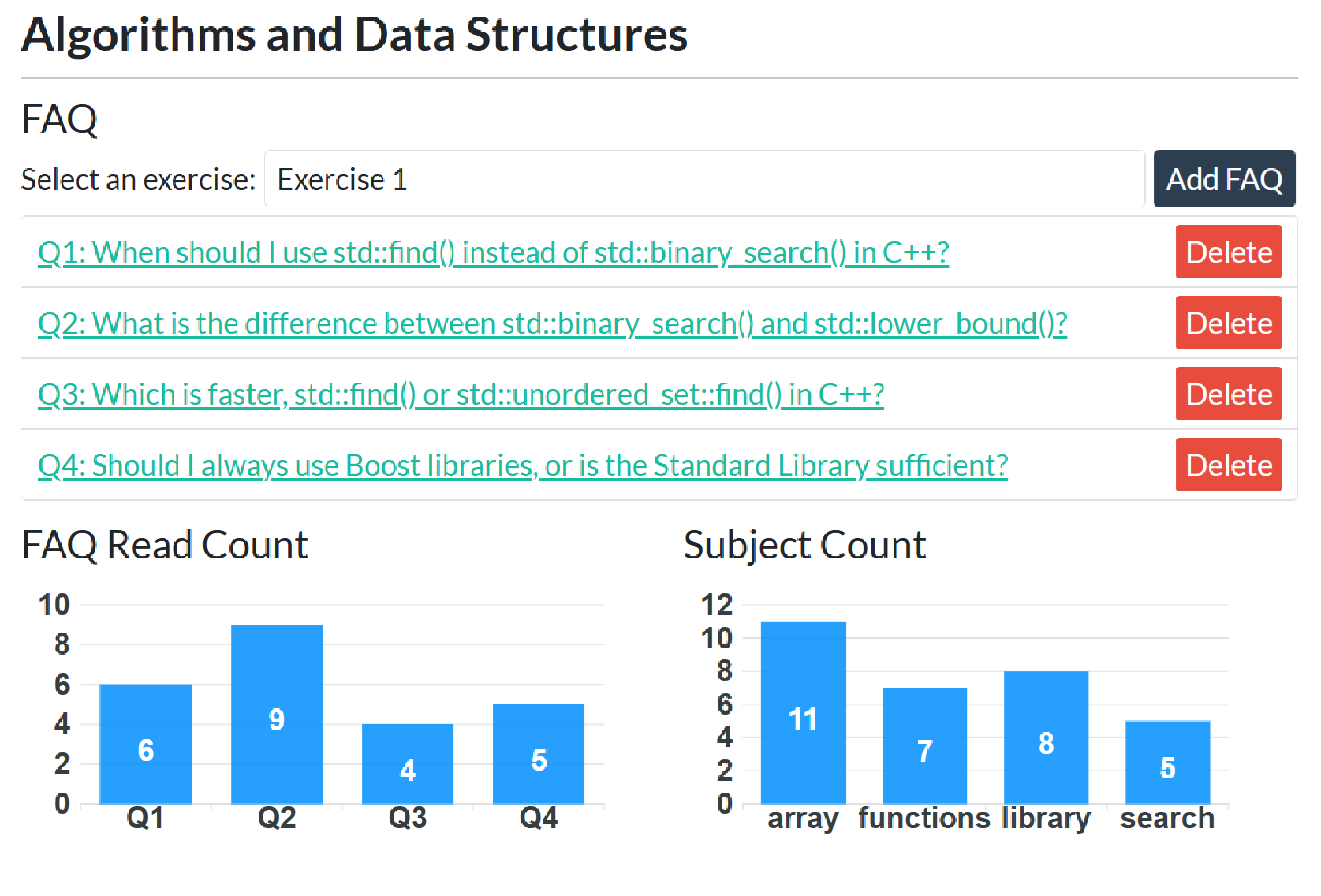}
    \caption{User interface for the teaching staff, with the dynamic FAQ and two visualizations. The teaching staff can edit the FAQ items and their content, or add an FAQ item manually. The visualizations show the frequency with which each FAQ item has been viewed and the frequency of topics in the questions from students to the LLM.}
    \label{fig:teacher-ui}
\end{figure}

\begin{figure}
    \centering
    \includegraphics[width=0.75\linewidth]{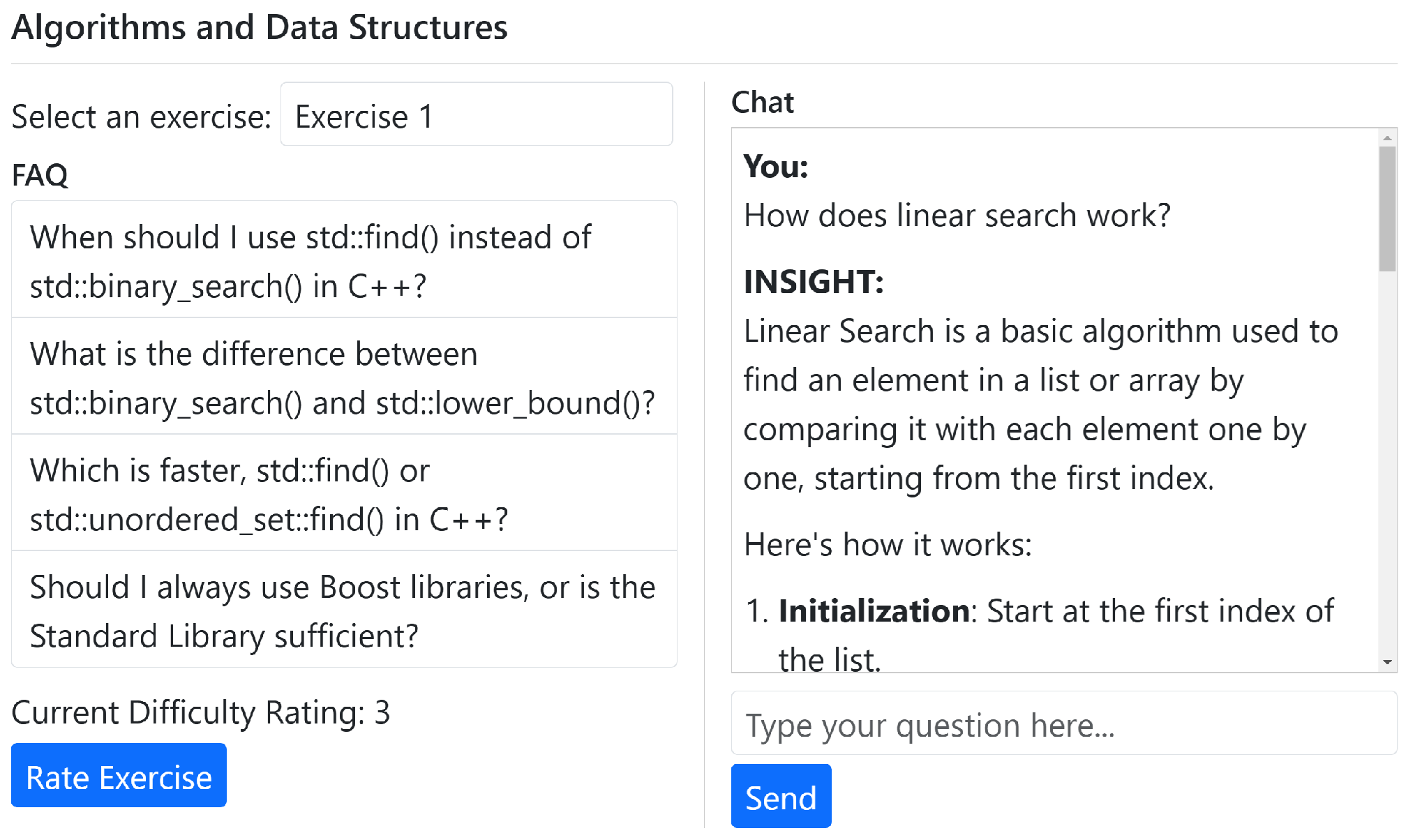}
    \caption{The student interface includes the FAQ and a chat window for the LLM. FAQ views are counted, optionally attaching the student's name with their explicit consent. Chat questions are analyzed to visualize question topics for teaching staff. Students can also rate exercise difficulty to provide explicit feedback.}
    \label{fig:student-ui}
\end{figure}

\paragraph{\textbf{User interface.}}
The teaching staff's user interface, illustrated in \cref{fig:teacher-ui}, has two main components: a dynamic FAQ and two visualizations. The dynamic FAQ displays frequently asked student questions, and staff can edit the LLM-generated answers to ensure accuracy. The two graphs visualize FAQ view frequencies and the frequency of course topics in student queries to the LLM. The student interface, illustrated in \cref{fig:student-ui}, also includes the FAQ and an additional chat interface for the LLM.

\section{Discussion}
\codename presents an approach to integrating AI in education to enhance student-teacher interaction while maintaining transparency in data collection and ensuring users retain control over their data. Its modular design allows for flexible integration in various courses, while its AI-driven FAQ and keyword analysis help identify common student challenges. This knowledge can help teaching staff in personalizing face-to-face support for groups or individual students, depending on whether the individual student wants to share their data.

Despite these advantages, potential risks must be considered, including over-reliance on AI and the accuracy of generated responses. While the LLM's capabilities and limitations with regard to answering students' questions are not the focus of this work, advanced LLM prompting strategies, such as Chain-of-Thought~\cite{wei_chain--thought_2022}, and the use of a larger LLM, could be worthwhile improvements.

Importantly, AI tools are now ubiquitous, and it is difficult, if not impossible, to control how students use them. Responsible AI use relies not only on system design but also on human practices like dialogue, trust-building, and setting shared expectations, elements that must be cultivated outside the boundaries of any technical system. Because of this, \codename is not designed to impose hard constraints but instead tries to adopt an approach that encourages open conversations between students and teaching staff about how and when to use AI responsibly, using the FAQ system as a guide.

Furthermore, as students are aware that \codename monitors their questions, they may ask questions to another LLM outside of \codename. This can create a biased view for the teaching staff and skew their perception of students' knowledge. To encourage the adoption of \codename, students have to notice a positive impact on their learning experience using this tool, and thus, teaching staff need to actively use the new insights to improve student-teacher interactions. The teaching staff also has to communicate to the students when to use \codename, as it is complementary to the staff and not a replacement. Certain questions are best answered by the teaching staff through direct, in-person interaction with students. Also note that \codename fits within Hasselt University's approach of interactive classes with teaching assistants, but may need to be revised to accommodate other class settings.

Finally, we want to emphasize that \codename is not intended to prevent students from using LLMs to obtain direct answers. Instead, it creates an environment where students can use these tools with privacy safeguards in place, and where the resulting interaction data can be meaningfully leveraged by both students and teaching staff to enhance learning. While we recognize the value of scaffolding-based tools that guide students toward discovering answers themselves, \codename addresses a different but equally important need: providing a safe and transparent space for students who primarily seek answers, ensuring that their use of LLMs still contributes to their learning process and informs face-to-face support.

The next step in our research is to empirically validate the usefulness of \codename. To achieve this, we will first perform a cognitive walkthrough with new members of teaching staff who have expressed interest in piloting the tool. Afterward, we will introduce \codename as an assistive tool in various courses at Hasselt University. Future research opportunities lie in integrating knowledge tracing to support advanced personalization while providing more granular data on the uptake of specific skills to the teaching staff. Furthermore, future work could use the collected data to provide adaptive learning and adjust content dynamically based on student progress and learning styles, aiming to offer a more interactive and inclusive learning experience.

\section{Conclusion}
\codename is our attempt to exploit the potential of AI-driven educational tools to enhance both student learning and teaching staff support. \codename's goal is to help teaching staff identify knowledge gaps, personalize face-to-face support, and maintain student engagement by providing a modular, data-driven approach to analyzing student interactions. Its emphasis on privacy and transparency ensures that students can benefit from AI assistance without blindly giving away their data. In the next steps of our research, we will test \codename's effectiveness in real-world courses to validate our approach.

\begin{acknowledgments}
This work was supported by the Special Research Fund (BOF) of Hasselt University (BOF24OWB28 and BOF23OWB31). This research was made possible with support from the MAXVR-INFRA project, a scalable and flexible infrastructure that facilitates the transition to digital-physical work environments. The MAXVR-INFRA project is funded by the European Union - NextGenerationEU and the Flemish Government.
\end{acknowledgments}

\section*{Declaration on Generative AI}
During the preparation of this work, the authors used ChatGPT and Grammarly in order to: Improve writing style, Grammar and spelling check. After using these tool(s)/service(s), the authors reviewed and edited the content as needed and take full responsibility for the publication’s content. 

\bibliography{references}

\end{document}